\documentclass{ijcaArticle}
\usepackage{amsmath}
\usepackage{amssymb}
\ijcaVolume{VV}
\ijcaNumber{N}
\ijcaYear{YYYY}
\ijcaMonth{Month}

\ijcaVolume{*}
\ijcaNumber{*}
\ijcaYear{2012}
\ijcaMonth{--------}
\begin{document}
\title{ Efficient Multi Secret Sharing with Generalized Access Structures } 
\author{ 
   \large Binu V P \\[-3pt]
   \normalsize Research Scholar  \\[-3pt]
    \normalsize Dept Of Computer Applications \\[-3pt]
    \normalsize CUSAT \\[-3pt]
    \normalsize	binuvp@gmail.com \\[-3pt]
  \and
   \large Sreekumar A \\[-3pt]
   \normalsize Associate Professor  \\[-3pt]
    \normalsize Dept Of Computer Applications \\[-3pt]
    \normalsize CUSAT \\[-3pt]
    \normalsize	askcusat@gmail.com \\[-3pt]
}

\terms{secret sharing, multi secret sharing}
\keywords{access structure, dynamic participants, cheating detection}
\maketitle
\begin{abstract} 
  Multi secret sharing is an extension of secret sharing technique where several secrets are shared between the participants, each according to a specified access structure.The secrets can be reconstructed according to the access structure by participants using their private shares.Each participant has to hold a single share, additional information are made available in a public bulletin board.The scheme is computationally efficient and also each participant can verify the shares of the other participants and also the reconstructed secret.The scheme doesn't need any secure channel also.
\end{abstract}
\section{INTRODUCTION}
Secret sharing schemes are important tool used in security protocols.Originally motivated by the problem of secure key storage by Shamir \cite{shamir1979}, secret sharing schemes have found numerous other applications in cryptography and distributed computing.Threshold cryptography \cite{desmedt1992shared}, access control \cite{naor1998access}, secure multi party computation \cite{ben1988completeness} \cite{chaum1988multiparty} \cite{cramer2000general}, attribute based encryption \cite{goyal2006attribute} \cite{bethencourt2007ciphertext}, generalized oblivious transfer \cite{tassa2011generalized}   \cite{shankar2008alternative}, visual cryptography  \cite{naor1995visual} etc., are the significant areas where the secret sharing techniques are effectively utilized.

In secret sharing, the secret is divided among $n$ participants in such a way that only designated subset of participants can recover the secret, but any subset of participants which is not a designated set cannot recover the secret.A set of participants who can recover the secret is called an \textit{access structure} or \textit{authorized set}, and a set of participants which is not an authorized set is called an \textit{unauthorized set} or \textit{forbidden set}.
The following are the two fundamental requirements of any secret sharing scheme.
\begin{itemize}
\item \textbf{Recoverability:}Authorized subset of participants should be able to recover the secret by pooling their shares.
\item \textbf{Privacy:}Unauthorized subset of participants should not learn any information about the secret.
\end{itemize}

Let $\mathcal{P}=\{P_i|i=1,2,\ldots,n\}$ be the set of participants and the secret be $K$ .The set of all secret is represented by $\mathcal{K}$.The set of all shares $S_1,S_2,\ldots,S_n$ is represented by $\mathcal{S}$.The participants set is partitioned into two classes.
\begin{enumerate}
\item The class of authorized sets $\Gamma$ is called the \textit{access structure.}
\item The class of unauthorized sets $\Gamma^c=2^\mathcal{P}\setminus \Gamma$
\end{enumerate}

Let as assume that $\mathcal{P},\mathcal{K},\mathcal{S}$ are all finite sets and there is a probability distribution on $\mathcal{K}$ and $\mathcal{S}$. $H(\mathcal{K})$ and $H(\mathcal{S})$ are used to denote the entropy of $\mathcal{K}$ and $\mathcal{S}$ respectively.

In a secret sharing scheme there is a special participant called \textit{Dealer} $\mathcal{D} \notin \mathcal{P}$, who is trusted by everyone. The dealer chooses a secret $K \in \mathcal{K}$ and the shares $S_1,S_2,\ldots,S_n$ corresponding to the secret is generated.The shares are then distributed privately to the participants through a secure channel.
In the secret reconstruction phase, participants of an access set pool their shares together and recover the secret.Alternatively participants could give their shares to a combiner to perform the computation for them.If an unauthorized set of participants pool their shares they cannot recover the secret.Thus a secret sharing scheme for the access structure $\Gamma$ is the collection of two algorithms:\\

\textbf{Distribution Algorithm}:This algorithm has to be run in a secure environment by a trustworthy party called Dealer. The algorithm uses the function $f$ ,which for a given secret $K \in \mathcal{K}$ and a participant $P_i \in \mathcal{P}$, assigns a set of shares from the set $\mathcal{S}$ that is $f(K,P_i)=S_i \subseteq \mathcal{S}$ for $i=1,\ldots,n$.$$f:\qquad \mathcal{K} \times \mathcal{P} \implies 2^\mathcal{S}$$
\textbf{Recovery Algorithm}:This algorithm has to be executed collectively by cooperating participants or by the combiner ,which can be considered as a process embedded in a tamper proof module and all participants have access to it.The combiner outputs the generated result via secure channels to cooperating participants.The combiner applies the function $$g:\mathcal{S}^t \implies \mathcal{K}$$, to  calculate the secret.For any authorized set of participants $g(S_1,\ldots,S_t)=K$ if ${P_1,\ldots,P_t} \subseteq \Gamma$.If the group of participant belongs to an unauthorized set, the combiner fails to compute the secret.

A secret sharing scheme is called perfect if for all sets $B$, $ B \subset \mathcal{P}$ and $B \notin \Gamma$, if participants in $B$ pool their shares together they cannot reduce their uncertainty about $K$. That is, $H(K)=H(K\mid\mathcal{S}_B)$,where $\mathcal{S}_B$ denote the collection of shares of the participants in $B$.It is known that for a perfect secret sharing scheme $H(S_i) \geq H(K)$.If $H(S_i) = H(K)$ then the secret sharing scheme is called ideal.

An access structure $\Gamma_1$ is \textit{minimal} if $\Gamma_2 \subset \Gamma_1$ and $\Gamma_2 \in \Gamma$ implies that $\Gamma_2=\Gamma_1$.Only \textit{monotone access structure} is considered for the construction of the scheme in which $\Gamma_1 \in \Gamma$ and $\Gamma_1 \subset \Gamma_2$ implies $\Gamma_2 \in \Gamma$.The collection of minimal access sets uniquely determines the access structure.The access structure is the closure of the minimal access set.The access structure $\Gamma$ in terms of minimal access structure is represented by $\Gamma_{min}(\Gamma_0)$.

For an access structure $\Gamma$, the family of unauthorized sets $\Gamma^c=2^\mathcal{P} \setminus \Gamma$ has the property that given an unauthorized set $B \in \Gamma^c$ then any subset $C \subset B$ is also an unauthorized set.An immediate consequence of this property is that for any access structure $\Gamma$, the set of unauthorized sets can be uniquely determined by its \textit{maximal set}.$\Gamma^c_{max}$ is used to denote the representation of $\Gamma^c$ in terms of maximal set.

For all $B \in \Gamma$.If $|B| \ge t$ then the access structure corresponds to  a $(t,n)$ threshold scheme.In the $(t,n)$ threshold scheme $t$ or more participant can reconstruct the secret.

The section 2 gives an introduction of threshold secret sharing, section 3 explores the secret sharing technique based on generalized access structure,section 4 gives different multi secret sharing techniques.Section 5 deals with the proposed scheme.Section 6 is the analysis of the scheme.Conclusion and references are given in section 7 and 8.
\section{THRESHOLD SECRET SHARING }
Development of secret sharing scheme started as a solution to the problem of safeguarding cryptographic keys by distributing the key among $n$ participants and $t$ or more of the participants can recover it by pooling their shares.Thus the authorized set is any subset of participants containing more than $t$ members.This scheme is denoted as $(t,n)$ \textit{threshold scheme}.

The notion of a threshold secret sharing scheme is independently proposed by Shamir \cite{shamir1979} and Blakley \cite{blakley1979} in 1979.Since then much work has been put into the investigation of such schemes.Linear constructions were most efficient and widely used.A threshold secret sharing scheme is called perfect, if less than $t$ shares give no information about the secret.Shamir's scheme is perfect while Blakley's scheme is non perfect.Both the Blakley
and the Shamir constructions realize $t$-out-of-$n$ shared secret schemes.However, their constructions are fundamentally different.

Shamir's scheme is based on   polynomial interpolation over a finite field. It uses the fact that we can find a polynomial of degree $t-1$ given $t$ data points. A polynomial $f(x)=\sum_{i=0}^{t-1}a_ix^i$, with $a_0$ is set to the secret value and the coefficients $a_1$ to $a_{t-1}$ are assigned random values in the field.The value $f(i)$ is given to the user $i$ as a share. When $t$ out of $n$ users come together they can reconstruct the polynomial using Lagrange interpolation and hence obtain the secret.

Blakley's secret sharing scheme has a different approach and is based on hyperplane geometry. To implement a $(t,n)$ threshold scheme, each of the $n$ users is given a hyper-plane equation in a $t$ dimensional space over a finite field such that each hyperplane passes through a certain point. The intersection point of these hyperplanes is the secret. When $t$ users come together, they can solve the system of equations to find the secret.

McEliece and Sarwate \cite{mceliece1981sharing} made an observation that Shamir's scheme is closely related to Reed-Solomon codes \cite{reed1960polynomial}.The error correcting capability of this code can be translated into desirable secret sharing properties.
           
 Karnin et al \cite{karnin1983} realize threshold schemes using linear codes. Massey \cite{massey1993minimal} introduced the concept of minimal code words, and provided that the access structure of a secret sharing scheme based on a $[n,k]$ linear code is determined by the minimal codewords of the dual code.   
   
Number theoretic concepts are also introduced for threshold secret sharing scheme. The Mingotee scheme \cite{mignotte1983} is based on modulo arithmetic and \textit{Chinese Remainder Theorem (CRT)}. A special sequence of integers called Mingotte sequence is used here.  The shares are generated using this sequence. The secret is reconstructed by solving the set of congruence equation using CRT. The Mingotte's scheme is not perfect. A perfect scheme based on CRT is  proposed by Asmuth and Bloom \cite{asmuth1983}. They also uses a special sequence of pairwise coprime positive integers. 
       
Kothari\cite{kothari1985generalized} gave a generalized threshold scheme. A secret is represented by a scalar, and a linear variety is chosen to conceal the secret. A linear function known to all trustees is chosen and is fixed in the beginning, which is used to reveal the secret from the linear variety. The $n$ shadows are hyperplanes containing the liner variety. Moreover the hyper planes are chosen to satisfy the condition that, the intersection of less than $t$ of them results in a linear variety which projects uniformly over the scalar field by the linear function used for revealing the secret. The number $t$ is called the threshold. Thus as more shadows are known more information is revealed about the linear variety used to keep the secret, however no information is revealed until the threshold number of shadows are known. He had shown that Blakley's scheme and Karnin's scheme are equivalent and provided algorithms to convert one scheme to another. He also stated that the schemes are all specialization of generalized linear threshold scheme. Brickell\cite{brickell1989some} also give a generalized notion of Shamir and Blackley's schemes using vector spaces.

Researchers have investigated $(t, n)$ threshold secret sharing extensively.Threshold schemes that can handle more complex access structures have been described by Simmons \cite{simmons1992} like weighted threshold schemes, hierarchical scheme,compartmental secret sharing etc.They were found a wide range of useful applications.Sreekumar et al \cite{sreekumar2009secret} in 2009, developed threshold schemes based on Visual cryptography.
\section{GENERALIZED SECRET SHARING }
 In the previous section, we mentioned that any  $t$ of the $n$ participants should be able to determine the secret. A more general situation is to specify exactly which subsets of participants should be able to determine the secret and which subset should not.In this section we give the secret sharing constructions based on generalized access structure.

Shamir \cite{shamir1979} discussed the case of sharing a secret between the executives of a company such that the secret can be recovered by any three executives, or by any executive and any vice-president, or by the president alone. This is an example of  \textit{hierarchical secret sharing} scheme. The Shamir's solution for this case is based on an ordinary $(3,m)$ threshold secret sharing scheme. Thus, the president receives three shares, each vice-president receives two shares and, finally, every  executive receives a single share.

The above idea leads to the so-called weighted(or multiple shares based) threshold secret sharing schemes. In these schemes, the shares are pairwise disjoint sets of shares, provided by an ordinary threshold secret sharing scheme. Benaloh and Leichter have proven in \cite{benaloh1990generalized} that there are access structures that can not be realized using such scheme.

Several researchers address this problem and introduced secret sharing schemes realizing the general access structure.The most effecient and easy to implement scheme was Ito, Saito,Nishizeki's \cite{ito1989secret} construction.It is based on Shamir's scheme.The idea is to  distribute shares to each authorized set of participants using multiple assignment scheme, where more than one share is assigned to a participant if he belongs to more than one minimal authorized subset.

A simple scheme is mentioned by Beimel \cite{beimel2011secret}, in which the secret $S \in {0,1}$ and let $ \Gamma$ be any monotone access structure. The dealer shares the secret independently for each authorized set
$ B \in \Gamma $,where $B=\{P_{i1},\ldots,P_{il}\}$.     
The Dealer chooses $l-1$ random bits $r_{1},\ldots,r_{l-1}$.
Compute $r_{l}= S \oplus r_{1} \oplus r_{2} \oplus \cdots \oplus r_{l-1}$, and the Dealer distributes share $r_{j}$ to $P_{ij}$.
For each set $ B \in \mathcal {A}$, the random bits are chosen independently and each set in $\Gamma$ can reconstruct the secret by computing the exclusive-or of the bits given to the set.The unauthorized set cannot do so.
    
The disadvantage with multiple share assignment scheme is that the share size depends on the number of authorized set that contain $P_{j}$.A simple optimization is to share the secret $S$ only for minimal authorized sets.Still this scheme is inefficient for access structures in which the number of minimal set is big (  Eg:$(n/2,n)$ scheme  ).The share size grows exponentially in this case.
     
Benalohand Leichter \cite{benaloh1990generalized} developed a secret sharing scheme for an access structure based on monotone formula.This generalizes the multiple assignment scheme of Ito,Saito and Nishizeki \cite{ito1989secret}.The idea is to translate the monotone access structure into a monotone formula.Each variable in the formula is associated with a trustee in $\mathcal{P}$ and the value of the formula is \textit{true} if and only if the set of variables which are true corresponds to a subset of $\mathcal{P}$ which is in the access structure. This formula is then used as a template to describe how a secret is to be divided into shares.

Brickell \cite{brickell1991classification}developed some ideal schemes for generalized access structure using vector spaces.Stinson \cite{stinson1992explication} introduced a monotone circuit construction based on monotone formula and also the construction based on public distribution rules.Benaloh's scheme was generalized by Karchmer and Wigderson \cite{karchmer1993span},who showed that if an access structure can be described by a small monotone span program then it has an efficient scheme.
      
Cumulative schemes were first introduced by Ito et al \cite{ito1989secret} and then used by several authors to construct a general scheme for arbitrary access structures.Simmons \cite{simmons1992} proposed cumulative map, Jackson \cite{jackson1993cumulative} proposed a notion of cumulative array.Ghodosi et al \cite{ghodosi1998construction} introduced simpler and more efficient scheme and also introduced capabilities to detect cheaters. Generalized cumulative arrays in secret sharing is introduced by Long \cite{long2006generalised}.
\section{MULTI SECRET SHARING}
There are several situations in which more than one secret is to be shared among participants.As an example, consider the following situation, described by \cite{simmons1992} .There is a missile battery and not all of the missiles have the
same launch enable code. A scheme is to be devised which will allow  any selected subset of users to enable different launch code.The problem is to devise a scheme which will allow any one,or any selected subset, of the launch enable codes to be activated in this scheme.This problem could be trivially solved by realizing different secret sharing schemes, one for each of the launch enable codes, but this solution is clearly unacceptable since each participant should remember too much information. What is really needed is an algorithm such that the same pieces of private information could be used to recover different secrets. 

One common drawback of all secret  sharing scheme is that, they are one-time schemes.That is once a qualified group of participants reconstructs the secret $K$ by pooling their shares, both the secret $K$ and all the shares become known to everyone, and there is no further secret.In other words, each share kept by each participant can be used to reconstruct only one secret.

Karnin, Greene and Hellman \cite{karnin1983} in 1983 mentioned the  multiple secret sharing scheme where threshold number of users can reconstruct multiple secrets at the same time. Alternatively the scheme can be used to share a large secret by splitting it into smaller shares.Franklin et al \cite{franklin1992communication}, in 1992 used a technique in which the polynomial-based single secret sharing is replaced with a scheme where multiple secrets are kept hidden in a single polynomial.They also considered the case of dependent secrets in which the amount of information distributed to any participant is less than the information distributed with independent schemes.Both the schemes are not perfect. They are also one time threshold schemes. That is, the shares cannot be re used.

Blundo et al \cite{blundo1993efficient},in 1993 considered the case in which $m$ secrets are shared among participants in a single access structure $\Gamma$, in such a way that any qualified set of participants can reconstruct the secret.But any unqualified set of participants knowing the value of  number of secrets might determine some( possibly no)information on other secrets.Jackson et al \cite{jackson1994multisecret}, in 1994  considered the situation in which there is a secret $S_K$ associated with each $K$-subset of participants
and $S_K$ can be reconstructed by any group of $t$ participants in $K$
$(t\le K)$.That is each subset of $K$ participants is associated with a secret which is protected by a $( t , K$)-threshold access structure.
These schemes are called  multi-secret threshold schemes.They came up with a combinatorial model and an optimum threshold multi secret sharing scheme.Information theoretic model similar to threshold scheme is also proposed for multi-secret sharing.They have generalized and classified the multi-secret sharing scheme based on the following facts.
\begin{itemize}
\item{should all the secrets be available for potential reconstruction during the lifetime of the scheme, or should the access of secrets be further controlled by enabling the  reconstruction of a particular secret only after extra information has been broadcast to the participants.}
\item{whether the scheme can be used just once to enable the secrets or should the scheme be designed to enable multiple use. }
\item{If the scheme is used more than once then the reconstructed secret or shares of the participants is known to all other participants or it is known to only the authorized set.} 
\item{The access structure is  threshold or generalized in nature.}
\end{itemize}

In 1994 He and Dawson \cite{he1995multisecret} proposed the general implementation of  multistage secret sharing.The proposed scheme allows many secrets to be shared in such a way that all secrets can be reconstructed separately.The implementation uses Shamir's threshold scheme and assumes the existence of a one way function which is hard to invert.The public shift technique is used here.A $t-1$ degree polynomial $f(x)$ is constructed first, as in Shamir's scheme.The public shift values are $d_i=z_i-y_i$.where $z_i=f(x_i)$.The $y_i$'s are send to the participants secretly.For sharing the next secret $h(y_i)$ is used, where $h$ is the one way function.The secrets are reconstructed in particular order, stage by stage and also this scheme needs $kn$ public values corresponds to the $k$ secrets.The advantage is that each participant has to keep only one secret element and is of the same size as any shared secret.In 1995 Harn \cite{harn1995efficient} shows an alternative implementation of multi stage secret sharing which requires only $k(n-t)$ public values.The implementation become very attractive, especially when the threshold value $t$ is very close to the number of participants $n$.That is for multistage $(n,n)$ secret sharing.In this scheme an $(n-1)$ degree polynomial $f(x)$ is evaluated at $(n-t)$ points and are made public.Any $t$ participants can combine their shares with the $(n-t)$ public shares to interpolate the degree $(n-1)$ polynomial.Multiple secrets are shared with the help of one way function as in He and Dawson scheme.

The desirable properties of a particular scheme depends on both the requirements of the application and also the implementation.Several multi secret threshold schemes are developed by the research community.In the proposed scheme we considered a multi-secret sharing scheme, realizing general access structure.\\

 A computationally secure secret sharing scheme with general access structure, where all shares are as short as the secret is proposed by Christian Cachin \cite{cachin1995line} in 1995.The scheme also provides capability to share multiple secrets and to dynamically add participants on-line without having to redistribute new shares secretly to the current participants.These capabilities are achieved by storing additional authentic information in a publicly accessible place which is called a noticeboard or bulletin board.This information can be broadcast to the participants over a public channel.The protocol gains its security from any one-way function.Multi secret sharing in this scheme needs different one way functions.The shares are exposed during the reconstruction and hence cannot be reused.A distributed evaluation sub protocol is proposed by Goldreich et al \cite{goldreich1987play} using one way function, but this allows the secret to be reconstructed in a specified order.\\

 Pinch \cite{pinch1996line} in 1996 proposed a modified algorithm based on the intractability of the Diffie-Hellman problem, in which arbitrary number of secrets can be reconstructed without having to redistribute new shares.This scheme is multi use but the participant has to follow a sequence.Ghodosi et.al \cite{ghodosi1997prevent} showed that Pinch's scheme is vulnerable to cheating and they modified the scheme to include cheating prevention technique. Yeun et al.,\cite{yeun1998identify} proposed a modified version of the Pinch's protocol which identifies all cheaters, regardless of their number, improving on previous results by Pinch and Ghodosi et al.\\
 
  An efficient computationally secure on-line secret sharing scheme is proposed by Re-Junn Hwang and Chin-Chen Chang in \cite{hwang1998line} 1998 .In this each participant hold a single secret which is as short as the shared secret.They are selected by the participants itself so a secure channel is not required between the dealer and the participants.Participants can be added or deleted and secrets can be renewed with out modifying the secret share of the participants.The shares of the participants is kept hidden and hence can be used to recover multi secrets.The scheme is multi use unlike the one time use multi secret sharing scheme.\\
  In Pinch's scheme high computation over head is involved and also sequential reconstruction is used in the recovery phase.In 1999 Sun \cite{sun1999line} proposed a scheme having the advantages of lower computation overhead and parallel reconstruction in the secret recovery phase.The security of the scheme is only based on one-way function,not on any other intractable problem.
  In 2006 Pang et al \cite{pang2006efficient} ~\cite{pang2006secure} proposed  efficient and secure multi secret sharing with general access structures.The proposed scheme is a modification based on this scheme.
  An efficient, renewable, multi use, multi-secret sharing scheme for general access structure is proposed by Angsuman Das and Avishek Adhikari \cite{das2010efficient} in 2010 .The scheme is based on one way hash function and is computationally more efficient.Both the combiner and the participants can also verify the correctness of the information exchanged among themselves in this.
\section{PROPOSED SCHEME}
The scheme is based on Shamir and discrete logarithm problem. The shares are generated by the participants and hence there is no need for a secure channel between the dealer and the participant. The pseudoshares are send to the dealer and it is difficult to get the shares from the pseudo shares because of the complexity of the discrete logarithm problem.Shared secret, participants set and access structures can be changed dynamically without updating participants secret shares.The degree of the polynomial is only one,so the computational complexity is also less.
\subsection{initialization Phase}
Let $P={P_1,P_2,\ldots,P_n}$ be the set of participants.$K_1,K_2,\ldots ,K_k$ be the set of secrets to be shared according to the access structure $\Gamma_1,\Gamma_2,\ldots,\Gamma_k$, ~where ~ $\Gamma_i=\{\gamma_{i1},\gamma_{i2},\ldots,\gamma_{it}\}$ is the access structure corresponds to the secret $K_i$.\\

Select two large prime $p$ and $q$ and let $n=p \times q$.Randomly select an integer $g$ from $[\sqrt{n},n]$ such that $g \ne p \;\mbox{or}\; g \ne q$.Choose another prime $m$ larger than $n$. The dealer publishes $g,n,m$ on the public bulletin.Each participant randomly select an integer $s_i$ from $[2,n]$ as secret share and compute $ps_i=g^{s_i}~\mbox{mod}~n$.The pseudo shares $ps_i$ are send to the dealer,who will then publish them in the public bulletin board.
\subsection{Secret Sharing}
In this phase, the dealer share the secrets corresponds to each access structure by publishing the values in the bulletin board, which is used by the participants to later reconstruct the secret.\\
Dealer randomly select an integer $s0_i$ from $[2,n]$ such that $s0_i$ is relatively prime to $\phi(n)$ and compute $ps0_i=g^{s0_i} ~\mbox{mod}~n$ corresponds to each secret $K_i$.Find $h0_i$ such that $s0_i \times h0_i\equiv 1 ~ \mbox{mod}~\phi(n)$.

Select an integer $a$ from $[1,m-1]$ and construct a polynomial $f_i(x)=K_i+a \times x ~\mbox{mod}~m$.
Select $t$ distinct random integers from $d_{i1},d_{i2},\ldots ,d_{it}$ from $[1,m-1]$ to denote the $t$ qualified sets in $\Gamma_i$.Compute $f_i(1)$ and for each subset $\gamma_{ij}=\{P_{1j},P_{2j},\ldots ,P_{lj}\}$ compute 
\begin{multline*}
H_{ij}=f_i(d_{ij})\bigoplus ps_1^{s0_i}~\mbox{mod}~n\bigoplus ps_2^{s0_i}~\mbox{mod}~n\bigoplus \ldots \\
\bigoplus ps_l^{s0_i}~\mbox{mod}~n
\end{multline*}
The dealer then publish
 \begin{multline*}
ps0_i,h0_i,f_i(1),H_{i1},H_{i2},\ldots,H_{it},d_{i1},d_{i2},\ldots,d_{it}
\end{multline*}
 corresponds to each secret $K_i$ and the access structure $\Gamma_i$. 
 The dealer also publishes $F(K_i,d_ij)$ corresponds to each secret and each authorized access set which can be used by the participant for verification after the secret recovery,where $F$ is a two variable one way function.
 \subsection{Secret Reconstruction}
 The participants from any authorized subset $(\Gamma_i)$ can reconstruct the secret $K_i$ as follows.
 
 If $\gamma_{ij}=\{P_{1ij},P_{2ij},\ldots,P_{lij}\}$ want to reconstruct $K_i$, each participant compute $x_{kij}=ps0_{i}^{s_k},k=1,\ldots,l$.These values are then delivered to the designated combiner. The combiner computes $$f_i(dij)\prime=H_{ij}\bigoplus x_{1ij}\bigoplus x_{2ij} \bigoplus \ldots \bigoplus x_{lij}$$.Using $f_i(1)$ , $f_i(d_{ij})\prime$ and $d_{ij}$'s, he can reconstruct the polynomial and hence recover the secret.
 \begin{align*}
 f_i(x) &= f_i(1) \times \frac{(x-d_{ij})}{1-d_{ij}}+f_i(d_{ij})\prime \times \frac{(x-1)}{d_{ij}-1}\\
& =\frac{x \times f_i(1)-d_{ij} \times f_i(1)-x \times f(d_{ij})\prime + f(d_{ij})\prime}{1-d_{ij}}
 \end{align*}
 The shared secret $K_i=f_i(0)$.
 Each participant of the authorized set can exchange $x_{ij}$ with other participants in the group and each member can compute the secret individually.This doesn't need a specified combiner and it also avoids the transmission of secret from the combiner to the participant.Each participant can verify the given $x_{ij}$ by the other participants and also the recovered secret by using the public values.
\section{Analyses and Discussions}
In the proposed scheme, the degree of the used Lagrange polynomial $f(x)$ is only 1,and we can construct $f(x)$ very easily.The other operation is just XOR operation which can also be computed very efficiently.Each participant select his share and compute the pseudo share $ps_i=g^{s_i}~\mbox{mod}~q$.This avoids the computational quantity of the dealer.This also avoids the need for a secure channel.
The proposed scheme does not need special verification algorithm to check whether each participant cheats or not.In the secret reconstruction phase the combiner can check whether $x_{i}$ is a true share by checking $x_i^{h0_i}=ps_i~\mbox{mod}~m$.That is $x_i^{h0_i}={(ps0_i)^{s_i})}^{h0_i}=(g^{s0_ih0_i})^{s_i}=g^{s_i}=ps_i~\mbox{mod}~m$.Each participant can verify the secret after recovery by computing the two variable one way function $F(K_i,d_{ij})$ and compare the result  with the public value.\\

In the reconstruction phase, each participant $P_{ij}$ in $\gamma_{ij}$ only provides a public value $x_{ij}$ and he does not have to reveal the secret share $s_i$.It is difficult to get the secret share from the public value $x_{ij}$ and  $ps_i$,because the discrete logarithm problem is hard to solve.The scheme is computationally secure.The shares can be reused and hence the scheme is a multi use scheme.The polynomial $f(x)$ can be reconstructed only if two pints are known.The point $(1,f(1))$ is known publicly but the  second point can be obtained only by the authorized set of participants using their private shares.\\

The important property of the proposed scheme is that the shared secret, the participant set and the access structure can be changed dynamically without updating any participant's secret shadow.In order to update the secret, the dealer need to create a new polynomial $f(x)$ and update  $f(1)$.If a new qualified set is to be added then $H_{t+1}$ and $d_{t+1}$ need to be added.New participant can be added accordingly.The public information corresponds to each modified authorized set must be recomputed and the old information must be updated in the public bulletin.Deleting a participant or deleting the authorized set containing the participant needs, deleting the public information corresponding to the access set.However for security reasons the secret also need to be updated.The scheme has following important properties.
\begin{enumerate}
\item The scheme can share multiple secrets, each with a specified access structure.
\item The participant has to hold only a single share in order to share multiple secrets.
\item The size of the share is as short as the secret
\item Participants select their secret shares and the dealer need not know the shares of the participants.This avoids the need of a secure channel.
\item The scheme is multi use ie; the participants can reuse the shares after a secret is recovered.
\item Each participant can verify the shares provided by the others in the recovery phase.
\item The dealer can modify the secret or add new secret with out modifying the participants secret shadow.
\item After the secret is recovered, the participants can verify the validity of the recovered secret.
\item The access structures can be dynamically modified.Only the public values need to be modified in this case also.
\end{enumerate}
\section{conclusion}
In this paper an efficient multi secret sharing scheme with a generalized access structure is proposed.The scheme is multi use and hence the shares can be reused by the participants.The participant select their secret shadows and the secret can be reconstructed by any participant in the authorized subset.No secure channel is required because the secrets or the secret shares are never send through the channel.The scheme is also verifiable because each participant can verify the shares of the other participants during the reconstruction phase and also the participants can verify the reconstructed secret.The shared secret, access structure or the participants set can be dynamically modified with out modifying the participants secret shadow.The scheme is also computationally efficient and can be implemented easily.
\bibliographystyle{ijcaArticle}
\bibliography{ss}
\end{document}